\documentstyle{mn}
\begin{document}


\title[Optical studies of XTE~J2123--058 -I] {Optical studies of the 
X-ray transient XTE J2123-058 -I. Photometry} 
\author[C.Zurita et al.]
{C.Zurita$^1$, J.Casares$^1$, T.Shahbaz$^2$, P.A.Charles$^2$, R.I.Hynes$^3$,
\and S.Shugarov$^4$, V.Goransky$^4$, E.P.Pavlenko$^5$, Y.Kuznetsova$^6$
\vspace{7mm} \\ 
1.Instituto de Astrof\'\i{}sica
de Canarias, 38200 La Laguna, Tenerife, Spain\\ 
2.Department of Astrophysics, Nuclear Physics
Laboratory, Keble Road, Oxford OX1 3RH,UK\\ 
3.The Open University, Walton Hall, Milton Keynes, MK7 6AA, UK\\
4.Sternberg State Astronomical Institute, 13 Universitetskiy
Prospect Moscow 119899, Russia\\
5.Crimean Astrophysical Observatory, 334413, Nauchny, Crimea,
Ukraine and Isaac Newton Institute Chile, Crimean Branch\\
6.Kyiv University, Physical Faculty, 64 Volodymyrska Str.,
Kyiv 252017\\
} 
\maketitle


\begin{abstract} 
We present optical photometry of the X-ray transient
XTE J2123--058, obtained in July--October 1998. The light curves are strongly
modulated on the 5.95hrs orbital period, and exhibit dramatic changes in
amplitude and form during the decline. We used synthetic models which include the 
effect of partial eclipses and X-ray heating effects, to estimate the system 
parameters, and we constrain the binary inclination to be i=73$^{\circ}\pm4$. The 
model is successful in reproducing the light curves at different stages of the 
decay by requiring the accretion disc to become smaller and thinner by 30\%
as the system fades by 1.7 mags in the optical. From Aug 26 the system
reaches quiescence with a mean magnitude of
R=21.7$\pm$0.1 and our data are consistent with the optical variability being
dominated by the companion's ellipsoidal modulation. 
\end{abstract} 




\section{Introduction} 
Soft X-ray transients (SXTs) are a
subclass of low-mass X-ray binaries  (LMXBs) that are characterized by 
episodic X-ray outbursts (usually  lasting for several months), when the X-ray
luminosities can increase by as much as a factor of 10$^7$ (van Paradijs \&  
McClintock,
1995). The observed optical flux  is generated by X-ray re-processing in the accretion
disc and  the companion star. These outbursts recur on a time scale of decades,
but  in the interim the SXTs are in a state of quiescence  and the optical
emission is dominated by  the radiation of the faint companion star. This
offers the best  opportunity to analyze the properties of this star and obtain
dynamical  information which eventually enables us to constrain the nature of
the compact object. There are currently 12 SXTs with identified optical
counterparts,  with 8 dynamical black-holes and 3 confirmed neutron stars:
CenX-4,  Aql X-1 and 1608-522 (van Paradijs \& McClintock, 1995). 
In addition, there are a few neutron star binaries exhibiting X--ray on and off 
states ( EXO0748--676, 4U 2129+47 and SAX J1748.9--2021 ), although they are 
not classified as SXTs because they do not show the classic fast rise and slow exponential/linear 
decay.

The X-ray transient J2123--058 was discovered on 29 June 1998 by the  Rossi
X-Ray Time Explorer ({\it RXTE}) (Levine  et al. 1989)  reaching a
peak X-ray flux of 100 mCrabs (2-12 keV). Its high Galactic  latitude
(b=-36$^{\circ}$.2) is unusual among transients, an indication  that 
J2123--058
might be a member of the galactic halo population. The  optical counterpart 
was
promptly identified with a variable star of R=17.2  (Tomsick et al. 1998a),
which was only marginally visible on  a digitized U.K. Schmidt plate,
suggesting a preoutburst magnitude  R$\ge$20 (Zurita et al. 1998). Spectra
obtained early in the outburst  showed strong high-excitation lines of 
He\,{\sc ii} $\lambda$4686, C\,{\sc iii}/N\,{\sc iii} $\lambda$4640 and weak Balmer
emission embedded in broad absorptions  (Tomsick et al. 1998a, Hynes et al.
1999). These features are  frequently observed in SXTs during outburst (e.g.
Callanan  et al 1995) and persistent LMXBs (e.g. Augusteijn et al. 1998). 
Type-I (thermonuclear) bursts have been detected both in X-rays (Takeshima  
and
Strohmayer 1998) and optical (Tomsick et al. 1998b), a signature of a
neutron star in the binary. The outburst light curve exhibited regular  0.7 mag
deep triangular-shaped minima repeating every 6hrs (Casares et al.  1998,
Tomsick et al. 1998b), a strong indication of high inclination. This  provided
the first evidence for the system orbital period  (P= 5.957$\pm$0.003) which
was later confirmed by a radial velocity study of the He{\sc ii}
$\lambda$4686 emission line (Hynes et al. 1998). In addition, Ilovaisky and
Chevalier (1998) reported the presence of a 0.3  mag modulation with a period
of 7.2 days, probably caused by the precessing  disc. Since 26 Aug the system had
settled down to its quiescent state at  R$\simeq$21.7 (Zurita \& Casares
1998).

This paper presents the results of a comprehensive set of observations that  have
led to detailed optical light curves from outburst through the decay  into
quiescence. Our spectroscopy will be the subject of a second paper  (Hynes et
al. 2000).

\vskip 10mm

\begin{tabbing}
{\bf Table 1} \ Log of observations.
\end{tabbing}


\begin{tabular}{lcrc}
\hline 
\hline
{\em Date} & {\em HJD$^{(*)}$} & {\em Exp/Filter} & {\em Telescope}\\ 
\hline 
{\it Jul02 98}   & -3  & 1xR               & IAC80 0.8m$^{(1)}$\\
{\it Jul04 ''}   & -1  & 1xB,1xV,1xR       & IAC80 0.8m\\
{\it Jul06 ''}   & 01  & 80xR              & OGS 1m$^{(2)}$\\
{\it Jul07 ''}   & 02  & 79xR              & OGS 1m    \\
{\it Jul08 ''}   & 03  & 20xR              & OGS 1m    \\
{\it Jul09 ''}   & 04  & 33xR              & OGS 1m    \\
{\it Jul10 ''}   & 05  & 20xR              & OGS 1m    \\
{\it Jul12 ''}   & 07  & 42xR              & OGS 1m    \\
{\it Jul13 ''}   & 08  & 24xR              & OGS 1m    \\
{\it Jul14 ''}   & 09  & 2xV               & M. Canopus 1m$^{(3)}$\\
{\it Jul16 ''}   & 11  & 2xV               & M. Canopus 1m\\
{\it Jul18 ''}   & 13  & 1xB,1xV,1xR       & M. Canopus 1m\\
{\it Jul19 ''}   & 14  & 40xBV             & Crimean 0.5m$^{(4)}$\\
              &     & 36xR              & Crimean 1.25 m$^{(5)}$\\
              &     & 2xB,2xV,2xR       & M. Canopus 1m\\
{\it Jul20 ''}   & 15  & 42xBV             & Crimean 0.5m\\
              &     & 74xR              & Crimean 1.25m\\
              &     & 1xB,5xV,1xR       & M. Canopus 1m\\
{\it Jul21 ''}   & 16  & 36xBV             & Crimean 0.5m\\
              &     & 94xR              & Crimean 1.25m\\
{\it Jul22 ''}   & 17  & 25xBV             & Crimean 0.5m\\
              &     & 43xR              & Crimean 1.25m\\
{\it Jul23 ''}   & 18  & 48xBV             & Crimean 0.5m\\
              &     & 45xR              & Crimean 1.25m\\
{\it Jul24 ''}   & 19  & 61xR              & Kryonerion 1.2m$^{(6)}$\\
              &     & 55xBV             & Crimean 0.5m\\
{\it Jul26 ''}   & 21  & 70xR              & OGS 1m\\
              &     & 46xBV             & Crimean 0.5m\\
{\it Jul27 ''}   & 22  & 77xR              & OGS 1m\\
              &     & 33xBV             & Crimean 0.5m\\
{\it Jul28 ''}   & 23  & 77xR              & OGS 1m\\
              &     & 40xBV             & Crimean 0.5m\\
{\it Jul29 ''}   & 24  & 74xR              & OGS 1m\\
              &     & 49xBV             & Crimean 0.5m\\
{\it Jul30 ''}   & 25  & 81xR              & OGS 1m\\
              &     & 33xBV             & Crimean 0.5m\\
{\it Jul31 ''}   & 26  & 40xBV             & Crimean 0.5m\\
\hline 
\hline
\end{tabular}
\\

\begin{tabbing}
\end{tabbing}
\begin{tabular}{lcrc}
\hline 
\hline
{\em Date} & {\em HJD$^{(*)}$} & {\em Exp/Filter} & {\em Telescope}\\ 
\hline 
{\it Aug01 98}   & 27  & 23xBV             & Crimean 0.5m\\
              &     & 4xR               & Crimean 0.38m\\
{\it Aug02 ''}   & 28  & 1xR               & IAC80 0.8m\\
{\it Aug03 ''}   & 29  & 1xR               & IAC80 0.8m\\
{\it Aug04 ''}   & 30  & 2xR               & Crimean 0.38m\\
              &     & 1xR               & IAC80 0.8m\\
{\it Aug05 ''}   & 31  & 2xR               & Crimean 0.38m\\
{\it Aug12 ''}   & 38  & 1xR               & IAC80 0.8m\\
{\it Aug15 ''}   & 41  & 12xR              & IAC80 0.8m\\
{\it Aug16 ''}   & 42  & 26xR              & IAC80 0.8m\\
{\it Aug17 ''}   & 43  & 8xR               & IAC80 0.8m\\
{\it Aug19 ''}   & 45  & 3xV,1xR           & M. Canopus 1m\\
{\it Aug26 ''}   & 52  & 18xR              & OGS 1m\\
              &     & 1xV               & IAC80 0.8m\\
{\it Aug27 ''}   & 53  & 14xR              & OGS 1m\\
{\it Sep02 ''}   & 59  & 4xR               & IAC80 0.8m\\  
{\it Sep23 ''}   & 80  & 11xR              & OGS 1m\\
{\it Sep24 ''}   & 81  & 8xR               & OGS 1m\\
{\it Sep25 ''}   & 82  & 5xR               & OGS 1m\\
{\it Sep26 ''}   & 83  & 5xR               & OGS 1m\\
{\it Jun21 99}   &351  & 1xR,1xV,1xI    & JKT 1m$^{(7)}$\\
       &     &                  &       \\
\hline 
\hline

\end{tabular}
\vspace{5pt}
\footnotesize{$^*$HJD--2451000}\\
\footnotesize{$^1$IAC80--80cm Telescope in the Observatorio del Teide (Tenerife).}\\
\footnotesize{$^2$ 1 m Optical Ground Station in the Observatorio del Teide (Tenerife).}\\
\footnotesize{$^3$ Mount Canopus 1m telescope at Tasmania.}\\
\footnotesize{$^4$ 0.5m telescope at Crimea}\\
\footnotesize{$^5$ 1.25m reflector of SAI Crimean Station.}\\
\footnotesize{$^6$ 1.2m telescope at the Kryonerion Astronomical station of the 
National Observatory of Athens at Kryonerio of Korinthia.}\\
\footnotesize{$^7$ 1m Jacobus Kapteyn Telescope in the Roque de los Muchachos Observatory (La Palma).
\normalsize


\noindent
\section{Observations and Data Reduction}
We observed J2123-058 during the period 1998 July--September with the IAC80 and
1m Optical Ground Station (OGS) in the Observatorio del Teide (Tenerife), the
1.25m SAI Station and 0.5m and 0.38m telescopes at Crimea, the Mount Canopus 1m telescope
at Tasmania and the 1.2m telescope at the Kryonerion Astronomical station of
the National Observatory of Athens at Kryonerio of Korinthia.
On the night of 21 of June of 1999, we obtained VRI images, using the 1m 
Jacobus Kapteyn Telescope in the Roque de los Muchachos Observatory (La Palma). The
 integration times
ranged from 1 to 40 min, depending on the telescope, atmospheric conditions and
the brightness of the target. The observing log is presented in Table 1. All images were
corrected for bias and flat-fielded in the standard  way.

We applied aperture photometry to our object and several nearby comparison
stars within the field of view, using {\sc iraf}. We selected four comparison
stars which were checked for variability  during each night, and
over the entire data set. We calibrated the data was using 9 standard
stars from several fields (Landolt 1992), from which we constructed
a colour dependent calibration. The Crimean data in R were calibrated
relative to the nearby comparison star from the USNO catalogue which was also in
to the previous calibrated data. White light photometry was also obtained in
Crimea using a 0.5m telescope equipped with a TV detector. The TV camera 
operates in the range $\lambda\lambda$3500--8000  with maximum sensitivity at
$\lambda$4000 and effective wavelength at $\lambda_{eff}$4851. Magnitudes were
measured relative to a nearby star from the USNO catalogue which was calibrated
at B and V wavelengths. 'Equivalent BV magnitudes', were calculated by interpolating the
fluxes in B and V to find the flux at the effective wavelength of the TV
detector. For further details see Pavlenko, Prokofieva \& Dolgushin (1989).

\section{Light Curves} 
\subsection{Long term light curve} 

Figure 1 compares the overall light curve of J2123--058 in optical
(VR bands)  and X-rays (2--12 keV). The long-term behaviour in X-rays shows a
classical {\it FRED} (fast-rise exponential-decay) morphology, with
characteristic  e-folding times of 2.4 d (rise) and 19 d (decay). These
time scales coincide  with the mean values of the distributions of SXTs (Chen,
Shrader \& Livio  1997). Twenty days after the peak of the outburst, the X-ray
intensity reaches a secondary maximum (of less than half of the peak
intensity). 
The secondary maximum is also suggested by our optical data.

We identify 3 different stages in the optical light curve: the outburst 
plateau (until $\sim$ 10 Aug), the decay phase ($\simeq$ 10--26 Aug) and 
quiescence (from $\simeq$ 26 Aug). In the plateau phase the object  brightness
decays at a moderately slow rate of $\simeq$ 0.03  mag$\cdot$day$^{-1}$
although a modulation is clearly visible in the nightly  mean magnitudes. The
time scale of this variability is consistent with the  7-d modulation attributed
to disc precession by Ilovaisky \& Chevalier  (1998). From $\sim$ 10 Aug the
optical light curve began an abrupt fall at  a rate of $\simeq$ 0.2
mag$\cdot$day$^{-1}$ before reaching quiescence on  26 Aug at R=21.7 (see
Figure 1).

Taking V(peak)$\simeq$17.2 and V(quiescent)$\simeq$22.9, we estimate a total 
outburst amplitude of 5.7 mags. Using the empirical relation 
$\Delta$V=14.36-7.6$\cdot\lg$P$_{orb}$(hr) (Shahbaz \& Kuulkers 1998), we would 
expect a total amplitude $\Delta$V$\simeq$8.4, 2.7 mags larger that what is 
observed. This difference can be explained by 3 effects: {\sc (1)} the outburst 
brightness of the  disc being reduced by a factor cos(i), since the disc would be 
foreshorted by the high binary inclination angle, {\sc (2)} we are also assuming 
that the quiescent flux is completely dominated by the companion star with no 
veiling from the accretion disc and {\sc (3)} if the secondary star is 
sufficiently evolved for it to be degenerate it would appear intrinsically 
fainter. We believe the discrepancy is due to a combination of these effects and
can only be resolved by obtaining optical spectroscopy in quiescence.

We see evidence for optical bursts both during outburst (29 July
with an  amplitude of 0.3 mag) and at the onset of quiescence (27 Aug 
and 2 Sept with amplitudes larger than 1 mag). The optical bursts 
observed during the ourburst decay is most probably caused by
re-processing of X-ray bursts. 
However, the origin of the optical bursts observed during the onset 
of quiesence is more puzzling, given that the source had almost reached 
its quiescent (low luminosity) X-ray state. Note that no simultaneous X-ray 
observations exist, which might shed light on the origin of these 
optical flashes observed during quiescence.

Our V and 'BV' (Figure 1), show that the BV magnitudes drop faster than R.
Moreover the amplitudes in BV and R increase during the decay.
Figure 2 presents colour information of J2123-058 as a function of time.
Although we cannot directly compare the 'BV'--R colour with V--R it is clear
from the plot that the system redden as the outburst decays and the
secondary's contribution increases. 
In quiescence, we obtained an upper limit to the quiescent V magnitude through 
2x2400s images of J2123--058 with the IAC80 telescope. We also have a marginal 
detection in V on the night of 21 Jun 1999 using the JKT telescope. Our colours are 
consistent with the spectral type of a late--K main sequence star.

\subsection{The orbital light curve}

Representative light curves of J2123-058
in different stages of the  outburst cycle are presented in Figure 3. The data
have been folded  on the updated ephemeris given by Zurita \& Casares (1998): HJD
2451042.639(5)  + 0.24821(3) E. Note the dramatic changes in amplitude and
morphology  of the light curves as the outburst decays.

The July light curves
are flat topped with broad triangular minima. They have a full amplitude of
0.7 mag and are reminiscent of the 5.1 hr  period eclipsing transient EXO
0748--676, although its long term  behaviour is characterized by high and low
X-ray/optical states rather than  transient outbursts. We also find
similarities with the eclipsing LMXB  source 2A1822--371 whose complex optical
light curve has been  successfully modeled by X-ray heating and partial
occultation of a  thick non-axisymmetric accretion disc and a faint companion
(Mason et al. 1980). Asymmetries in the eclipse minima, are also seen in
J2123-058 on individual nights and also in the curves by Tomsick et
al. 1998.

On the other hand, the light curve of Aug 16 is almost
sinusoidal and  shows a peak-to-peak amplitude of $\sim$1.4 mag. It resembles
the outburst  light curve of the 5.2 hr neutron star system 4U 2129+47, where
the large  amplitude has been attributed to X-ray reprocessed radiation from
the heated  face of the optical star (Thorstensen et al.1979). Note, however, 
that at the onset of the fast optical decay ($\simeq$10 Aug) the  X-ray flux
had already dropped by a factor $\ge$ 10 with respect to the  outburst peak. Note
also the presence of two narrow 0.2 mag dips  at phases 0 and 0.5 which suggest
the presence of grazing eclipses.

Finally, the quiescent light curve
displays a characteristic ellipsoidal  modulation from the secondary star: a
double-humped variation with a full  amplitude of $\sim$0.4 mag. This
light curve was produced by phase binning the entire quiescent data
(from Aug 26 on) after detrending the night-to-night variability using a
linear fit. 


\section{Distance estimate}
In the context of King \& Ritter's (1998), the X-ray exponential decay 
seen in J2123-058 indicates that irradiation is strong enough to ionize the 
entire accretion disc. Also, a secondary maximum is expected one 
irradiated-state viscous time after the onset of the outburst and it can be 
used to calibrate the peak X-ray luminosity and hence the distance to the 
source $D_{\rm kpc}$ through 
$$D_{\rm kpc}= 4.3 \times 10^{-5} t_{s}^{3/2} \eta^{1/2} f^{1/2} 
F_{p}^{-1/2} \tau_{d}^{-1/2}$$
\noindent
where $F_p$ is the peak X-ray flux, $t_s$ the time of the secondary 
maximum after the peak of the outburst in days, $\tau_d$ the e-folding time 
of the decay in days, $\eta$ the radiation efficiency parameter and 
$f$ the ratio of the disc mass at the start of the outburst to the 
maximum possible mass (Shahbaz, Charles \& King 1998). In our case, 
$\tau_d$=19 d, $t_{s}\simeq$20 d and $F_p$ can be estimated from the XTE 
count rate (6 counts s$^{-1}$ in the energy range 2-12 keV) which corresponds to 
1.84 $\times$ 10$^{-9}$ erg cm$^{-2}$ s$^{-1}$. Assuming $\eta$=0.15 and 
$f$=0.5 we find $D_{\rm kpc}$ = 5.7.

Alternatively, we can estimate the distance to the source by comparing the 
quiescent magnitude with the absolute magnitude of a main sequence star 
which fits within the Roche lobe of a 6hr period orbit. 
Combining Paczynski's (1971) expression for the  averaged radius of a
Roche lobe with  Kepler's Third Law we get the  well-known relationship between
the secondary's mean density and the  orbital period: $\rho = 110/ P_{\rm
hr}^{2}$ (g cm$^{-3}$). Substituting  the orbital period of J2123--058 we obtain
$\rho$=3.1 g cm$^{-3}$  which corresponds to a K7V secondary star with mass 
$\simeq$0.6 
M$_{\odot}$ and absolute magnitude $M_{R}\simeq$ 7. The dereddened quiescent magnitude is 
R=21.4 (using A$_V$=0.37$\pm$0.15 as derived from the NaD1 line; see Hynes 
et al. 1998) which yields $D_{\rm kpc}$ = 7.7. Strictly speaking, this is 
a lower limit to the distance as we are neglecting any contribution by the 
accretion disc to the quiescent optical flux. However, note that the true 
distance is probably not too far off 8 kpc since our quiescence light curve 
does not show strong evidence for disc contamination (see section 5 and Figure 6). 
A conservative limit can therefore be provided by assuming a 50 percent 
contribution by the accretion disc to the continuum light. 
Allowing for 50 percent disc contamination (as 
observed in J0422+32, a black-hole transient with comparable orbital period; 
see Casares et al. 1995) we obtain R=22.2 for the companion star and hence 
 $D_{\rm kpc}$ = 10.8. Hereafter we will adopt 
$D_{\rm kpc} = 8 \pm 3$ which is consistent, at the lower end, with other 
estimates based on photospheric expansion models of the X-ray bursts 
(Homan et al 1999, Tomsick et al 1999). A spectral type determination of the 
companion star is essential to refine this distance estimate.    


\section{Modelling} 

In an attempt to interpret the different light
curves and derive the system parameters, we have used a model based on the
work by de Jong et al.  (1996, also see references included). The model assumes 
a flared accretion  disc of the form $h\propto r^{9/7}$ (with $h$
and $r$ the
disc height and radius respectively) and a Roche lobe filling secondary and
accounts for X-ray heating,  shadowing effects and mutual eclipses of the disc
and the secondary. The disc is assumed to radiate as a blackbody  with a
radial temperature  distribution calculated according to Vrtilek et al. (1990).
The intensity  distribution on the secondary star is computed using Kurucz
model  atmospheres. The albedo of the accretion disc and the companion star
are  fixed to 0.95 and 0.40 respectively, following the results of de Jong et
al  (1996).  The model parameters are the binary inclination (i), mass  ratio
(q=M$_{1}$/M$_{2}$), the accretion disc radius (R$_{\rm disc}$) defined as a
fraction of the distance to the inner Lagrangian point  (R$_{\rm L_1}$), the
flaring angle of the accretion disc ($\alpha$) and  the X-ray luminosity
(L$_{\rm x}$).

In order to model the outburst light curve we have averaged  our best quality
light curves of the plateau phase (26-30 July) in  29 phase bins.   
Using L$_{x}$=1.3$\times$~10$^{37}$ erg~s$^{-1}$ (for $D_{\rm kpc}$ = 8)  
and M$_{2}$=0.6 M$_{\odot}$ we performed a least-squares fit to the data. 
Our best fit solution gave a reduced chi squared $\chi_{\nu}^{2}$ of 1.36 
for $i$=76.0$\pm$1.0 degrees, R$_{\rm disc}$=0.75$^{+0.06}_{-0.03}$ R$_{\rm
L_1}$, $\alpha$=7.6$^{+1.0}_{-0.2}$ degrees and $q$=4.6$^{+0.5}_{-0.2}$. The 
uncertainties quoted are at the 99 percent confidence level and were obtained 
by grid searching the parameter of interest whilst optimizing the other model 
parameters. 
We have also rescaled the $\chi_{\nu}^{2}$ values so that the 
minimum $\chi_{\nu}^{2}$ is 1. The best model fit to the outburst data 
is shown in Figure 4 (solid line).
In order to examine the effects of changing M$_2$, we fitted the outburst data
with M$_2$=0.1M$_\odot$ (Figure 4 -- dotted line). 
We find that the derived parameters are the same, within
the errors: $\chi^{2}_{\nu}=1.5$, $i$=75.3$^{+0.8}_{-1.2}$ degrees, 
R$_{\rm disc}$=0.83$\pm$0.05 R$_{\rm {L_1}}$, $q$=3.4$\pm$0.4, 
$\alpha$=8.8$^{+0.8}_{-1.2}$ degrees. 
Regarding the decay stage, we have fitted the light curve of 16 Aug binned into
20 phase bins, when the X-ray
luminosity was an order of magnitude lower, L$_{x}$=1.3$\times$10$^{36}$
erg~s$^{-1}$. We fitted the decay stage data with this X-ray
luminosity and M$_{2}$=0.6M$_{\odot}$.
Our best fit solution gave a reduced $\chi_{\nu}$ of 1.51 
for $i$=72.0$\pm$3.0 degrees, R$_{\rm disc}$=0.56$\pm$0.06 R$_{\rm
L_1}$, $\alpha$=5.7$\pm$0.5 degrees and $q$=4.2 ($q$ could not be
constrained with this data set).
The best model fit to the decay stage data 
is shown in Figure 5. Again 99 per cent confidence levels are quoted.

In Figure 6, we show the quiescent light curve which exhibits the
characteristic 
ellipsoidal modulation of the secondary star: i.e. two equal maxima and two unequal 
minima. The minimun at phase 0.5 is expected to be deeper than the minimum at 
phase 0.0 because gravity darkening is more pronounced near the inner Lagrangian 
point $L_{1}$. This effect is important for high inclination
systems (see eg. Avni and Bahcall, 1974). In the figure we also show model plots 
using i=73$^{\circ}$, q=4.6 (values which are consistent with those derived 
from the decay and outburst data), and no X--ray heating. The solid and 
dotted lines show plots with zero and 50 percent disc contamination. The 
former model probably best describes the data.


\section{Discussion}
XTE J2123--058 is a remarkable neutron star binary.
Our optical light curve 
shows marked orbital modulations with dramatic variations as the outburst 
declines. These are very similar to the modulations observed in accretion disc
corona (ADC) sources of comparable orbital periods (e.g. 4U 2129+47 and 2A
1822--371; see eg. McClintock et al. 1980 and Mason et al. 1982) although none
of these are transient systems.

The X-ray light curve displays the classic SXT properties, namely a FRED
morphology with typical e-folding  rise and decay times and secondary maximum.
Furthermore, the ratio of  X-ray to optical luminosity [$\xi=B_{0} +2.5 \log
F_{\rm x} (\mu\rm{Jy})$]  is also in excellent agreement with the observed
distribution of LMXBs.  Taking B=17.28 (Tomsick et al. 1998a) and F$_{\rm x}
(2-12 keV) \simeq  100$ mCrab (Levine, Swank \& Smith 1998) at the outburst
peak and assuming  $A_{\rm v}=0.37$ (Hynes et al 1998) we obtain $\xi$=21.9,
whereas the  distribution peak of LMXBs gives $\xi$=21.8 $\pm$ 1.0 (see van
Paradijs \&  McClintock 1995). This result implies that, despite its high 
inclination, the X-ray source in J2123-058 is not hidden by the accretion 
disc (i.e. $\alpha\geq$90$^{\circ}$--i). This is consistent with the 
values our model favors for the  binary inclination (i=73$^{\circ}$) and 
disc flaring angle ($\alpha$=5$^{\circ}$.7 -- 7$^{\circ}$.6). 

The longterm evolution of the optical light curve can be compared to  those of
other SXTs (e.g. GRO J0422+32, A0620-00, N. Muscae 91).  They show a
slow linear decay followed by a steeper fall.  We find that J2123--058
also reproduces this behaviour, although  the total amplitude and 
time scales are a factor of $\sim$ 2 shorter 
(see e.g. Callanan et al 1995).

We have modeled our R-band light curves of J2123--058 at different stages of 
the outburst 
including the obscuration effects and X-ray  heating 
of the secondary star accretion disc.  This led us to constrain the system
inclination to i=73$^{\circ}\pm4$. We find encouraging the excellent  agreement
between the inclination values obtained for the two independent  light curves
(July 26-30 and 16 Aug) during which $L_{\rm x}$ has dropped by  one order of
magnitude. The light curves at the plateau phase (July) are
very similar to those of  EXO 0748--676 and 2A 1822--371   with an extended
depression of the luminosity  from phase $\sim$0.7 until the eclipse and a
steeper rise to maximum  (see Mason et al. 1980 and Schmidtke et al 1987).  Our
model fit indicates that the accretion disc is the dominant source of  light  
and the triangular shaped minima can be interpreted as  eclipses of the
accretion disc by the secondary star together with  the changing aspect of the
heated polar caps of the companion star.  
The dramatic changes observed in the light curves during decline, are 
triggered by large changes in the disc size and geometry. Our fit to the decay data 
(16 Aug) demands a thinner  and smaller accretion disc which implies a smaller 
fraction of the disc 
is X-ray heated. Conversely, the secondary star is more exposed to the X-ray
radiation and therefore  the total amplitude of the modulation increases by a
factor of 2 (to $\sim1.4$ mag). The resulting light curve has a sinusoidal-like 
shape and  is reminiscent of the LMXBs 4U 2129+47(=V1727 Cygni) and HZ Her. Our 
model fits implies a change of $\sim$30 percent in the disc size, as the system 
fades by 1.7 mags in the optical. The 
change in the radius of the disc size is what one expects. If angular
momentum is transported outwards in the disc through viscous processes,
then at outburst, since matter diffuses inwards, the angular momentum of
that matter has to be transfered to the outer parts of the disc, and the
radius of the disc is expected to expand. When the system is decaying, after the 
end of the mass transfer enhancement, the disc shrinks to its original radius
(Livio \& Verbunt 1988; Ichikawa
\& Osaki 1992). Observations of U Gem, Z Cha, OY Car, HT Car show that
the accretion discs are indeed larger in outburst than in quiescence
(Smak 1984b; O'Donohuge 1986; Harrop-Allin \& Warner 1996). Comparing our results 
with disc 
radius variation in U Gem (Smak 1984a), we find approximately the same rate ofdecrease.

\section{Acknowledgments} 

Part of this work is based on
observations made with the European Space Agency OGS telescope operated on the
island of Tenerife by the Instituto de Astrofisica de Canarias in the Spanish
Observatorio del Teide of the Instituto de Astrofisica de Canarias.

We are grateful to M. Serra--Ricart, D.Alcalde, A.Gomez and P. Rodriguez--Gil 
for performing some of the observations. We are thank A. Dapergolas, E.T. 
Harlaftis and D. Galloway for making their data available to us. JC 
acknowledges support by the Spanish Ministry of Science grant 1995-1132-02-01.We thank J. van Paradijs and the Amsterdam group for the use of their X--ray 
binary model.

\section{References}
Augusteijn T., van der Hooft F., de Jong J.A., van Kerkwijk M.H., van
Paradijs J., 1998, A\&A, 332, 561\\ 
Avni, Y., Bahcall, J.N., 1974, ApJ, 192L, 139\\ 
Boehkarev,N.G., Karitskaia E.A.,  Shakura N.I., 1975, SvAL, 1, 118\\
Callanan P.J., Garcia M.R., McClintock J.E., Zhao P., Remillard R.A.,
Bailyn C.D., Orosz J.A., Harmon B.A., Paciesas W.S., 1995, ApJ, 441,
786\\  
Casares J., Martin A.C, Charles P.A., Mart\'\i{}n E.L., Rebolo R., 
Harlaftis E.T., Castro-Tirado A.J., 1995, MNRAS, 276, L35\\
Casares J., Serra-Ricart M., Zurita C., Gomez A., Alcalde D., Charles
P., 1998, IAU Circ. 6971\\
Chen W., Shrader C.R., Livio M., 1997, ApJ, 491, 312\\ 
de Jong J.A., van Paradijs J., Augusteijn T., 1996, A\&A, 314,484\\ 
Harrop-Allin M.K., Warner B., 1996, MNRAS, 279, 219\\
Homan J., M\'endez M., Wijnands R., van der Klis M., van Paradijs J., 
astro-ph/9901161\\
Hynes R.I., Charles P.A., Haswell C.A., Casares J., Zurita C., 1999, to 
appear in Proc. 19th Texas Symposium on Relativistic Astrophysics and Cosmology, Eds. J. Paul, T. Montmerle and E. Aubourg, CD--ROM Edn. (astro--ph/9904325)\\
Hynes R.I., Charles P.A., Haswell C.A., Casares J., Serra-Ricart M., Zurita C
., 1998, IAU Circ. 6976\\
Ichikawa S., Osaki Y., 1992, PASP, 44, 151\\
Ilovaisky S.A., Chevalier C., IAU Circ. 6975\\ 
King A.R., Kolb U., Burderi L., 1996, ApJ, 464, L127\\
King A.R., Ritter H., 1998, MNRAS, 193, L42\\
Lampton M., Margon E., Bowyer S., 1976, ApJ, 208, 177\\ 
Landolt A.U., 1992, ApJ, 140, 340\\ 
Levine A.,Swank J., Smith E. and RXTE ASM \& PCA teams, 1998,  IAU Circ. 6955\\
Livio M., Verbunt F., 1988, MNRAS, 232, 1\\
McClintock J.E., Remillard R.A., Margon B., 1981, ApJ, 243, 900\\  
Mason K.O., Cordova F.A., 1982, ApJ, 262, 253\\ 
Mason  K.O., Seitzer P., Tuohy I.R., Hunt L.K., Middleditch J., Nelson J.E., White N.E., 1998, ApJ, 242, 109\\ 
O'Donoghue D., 1986, MNRAS, 220, 23\\
Parmar A.N., White N.E., Giommi P. and Gottwald M., 1985, ApJ, 308,199\\ 
Pavlenko E.P. , Prokofieva V.V., Dolgushin A.I., 1989, Pis'ma Astr.
Zh., 15, 611\\ 
Schmidtke P.C., Cowley A.P., 1987, ApJ, 93, 374\\ 
Shahbaz T., Kuulkers E., 1998, MNRAS, 295, L1\\
Shahbaz T., Charles P.A., King A.R., 1998, 301, 382\\
Smak J., 1984 AcA, 34, 93\\
Smak J., 1984 AcA, 34, 161\\
Takeshima T., Strohmayer T.E., 1998, IAC Circ. 6958\\ 
Thorstensen J., Charles P.A., Bowyer S., Briel U.G., Doxsey R.E., Griffiths, 
R.E., Schwartz, D.A., 1979, ApJ, 233, 57\\  
Tomsick J.A., Halpern J.P., Leighly K.M., Perlman E., 1998, IAC Circ. 6957\\ 
Tomsick J.A., Kemp J., Halpern J.P., Hurley-Keller D., 1998, IAU Circ. 6972\\ 
Tomsick J.A., Halpern J.P., Kemp J., Kaaret P., astro-ph/9903360\\
van Paradijs J., McClintock J.E., 1995, in: X-Ray Binaries,  eds. W.H.G.
Lewin, J. van Paradijs \& E.P.J. van den Heuvel (Cambridge: Cambridge University Press), p. 58\\  
Vrtilek S.D., Raymond J.C., Garcia M.R., Verbunt F., Hasinger G., K\"{u}rster M., 1990, A\&A, 235, 162\\
Zurita C., Casares J., 1998, IAU Circ. 7000\\ 
Zurita C., Casares J., Hynes, R.I., 1998, IAU Circ. 6993\\

\newpage
\onecolumn
\begin{figure} 
\begin{center}
\begin{picture}(150,290) 
\put(0,0){\includegraphics{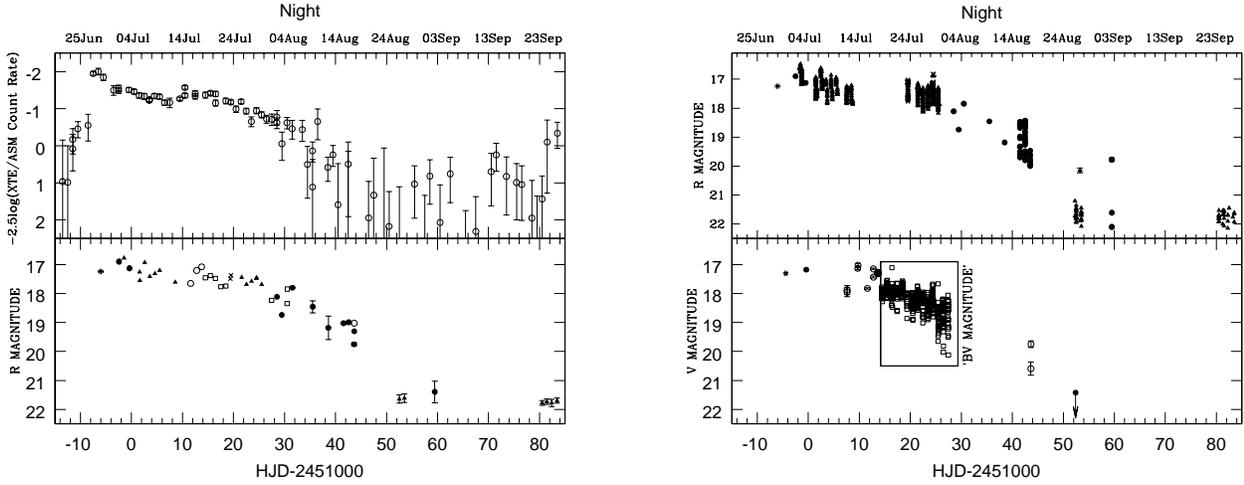}} 
\noindent 
\end{picture}
\caption[{\bf Figure 1}] { Left: Temporal evolution of J2123--058
plotted as 'X-ray magnitudes'(top) and R-band magnitudes averaged per day (bottom).
 Note that 1 Crab equals an {\sc asm} count rate of 75 counts per sec. The X--ray 
data are provided by the {\sc asm/rxte} teams at {\sc mit} and at the 
{\sc rxte sof} at {\sc Nasa}'s {\sc gsfc}.
Right: Temporal evolution plotted as R-band magnitudes (top) and V and 'BV' 
magnitudes (bottom).\\
 The optical data points have been obtained with the following telescopes/sites:
 {\sc OGS} (triangles),{\sc IAC80} (filled circles), Crimean telescopes (open 
squares), Kryoneri Astronomical Station (diagonal crosses) and Tasmania (open 
circles). Asterisks mark the magnitudes reported by Tomsick et al. (IAUC 6957).
\label{fig1}} 
\end{center}
\end{figure} 
\twocolumn

\begin{figure} 
\begin{center}
\begin{picture}(150,290) 
\put(0,0){\includegraphics{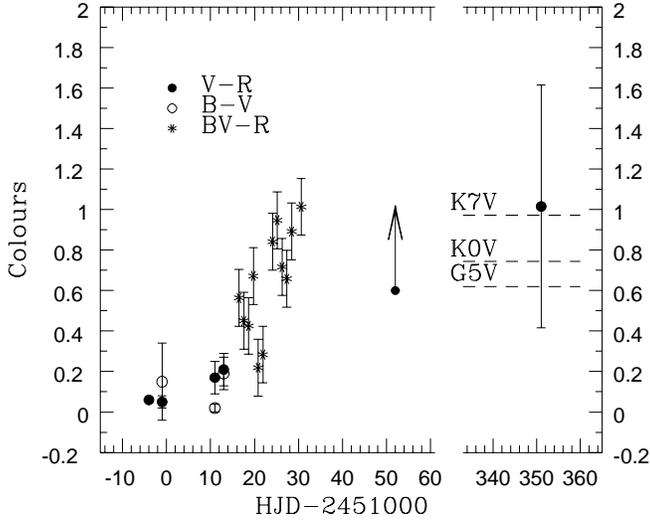}} 
\noindent
\end{picture}

\caption[{\bf Figure 2}] {(BV--R),(B--V) and (V--R) colours of J2123-058 as a
 function of time. Indices (B--V)
and (V--R) were obtained from single simultaneous points. (BV--R) were
calculated averaging magnitudes simultaneous within 3 min. Horizontal lines
mark (V--R) colours for different spectral types. 
\label{fig2}} 
\end{center}
\end{figure} 


\begin{figure} 
\begin{center}
\begin{picture}(150,275)
\put(0,0){\includegraphics{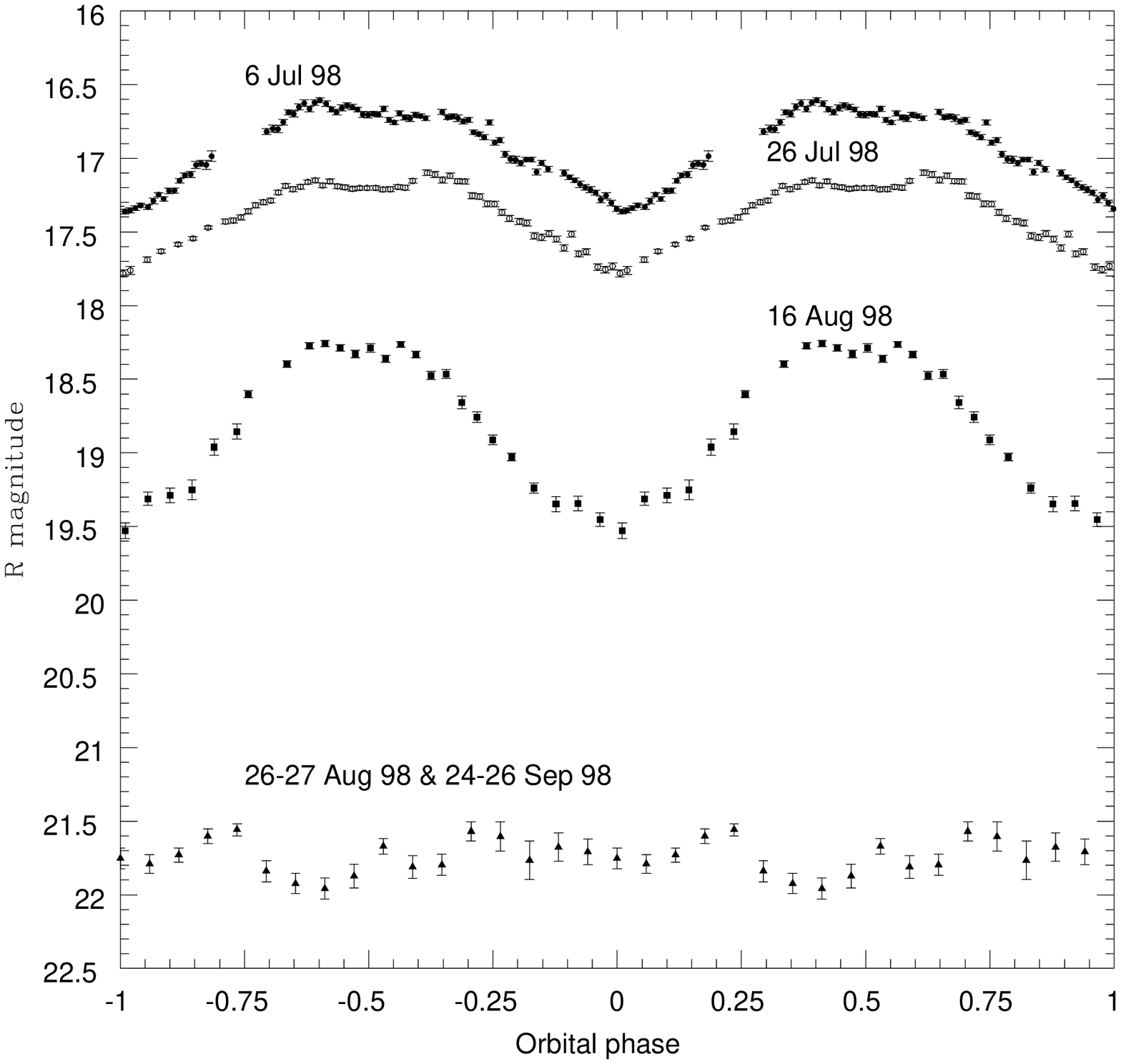}} 
\noindent 

\end{picture} 

\caption[{\bf Figure 3}]

{Optical light curves of J2123-058 taken during  July-Sep 1998, covering
outburst, decay and quiescence (bursts have been suppress in the averaged curves).
They show  spectacular  variations, evolving
from a strong triangular shape to sinusoidal and  double-hump modulations.
The light curves are folded on the 5.95 hr period and  are shown twice for clarity

\label{fig3}} 

\end{center} 

\end{figure} 



\begin{figure} 
\begin{center}

\begin{picture}(150,230) 
\put(0,0){\includegraphics{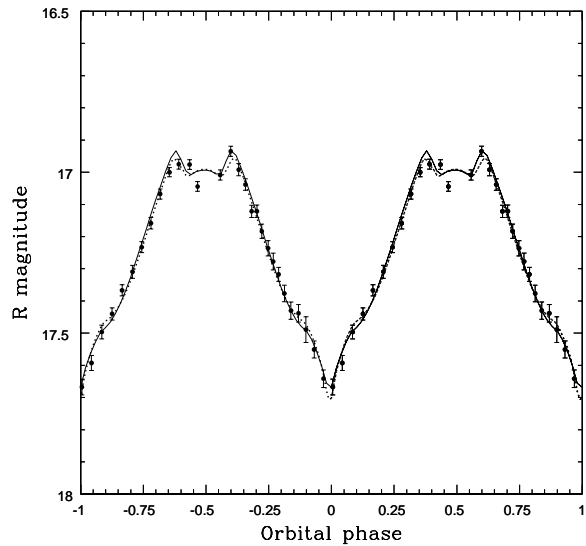}} 
\noindent 

\end{picture}

\caption[{\bf Figure 4}] {Outburst light curve (Jul 26--30) (points) and 
model fits (solid line) for L$_{x}$=1.3$\times$10$^{37}$erg~s$^{-1}$, 
i=76$^{\circ}$, q=4.6 (M$_{2}$=0.6M$_{\odot}$), R$_{disc}$=0.75 R$_{\rm L_1}$ 
and 
$\alpha$=7$^{\circ}$.6. 
The dotted line shows the best fit using M$_{2}$=0.1M$_{\odot}$ (see text).
\label{fig4}} 
\end{center} 
\end{figure} 

\begin{figure} 
\begin{center}
\begin{picture}(150,230) 
\put(0,0){\includegraphics{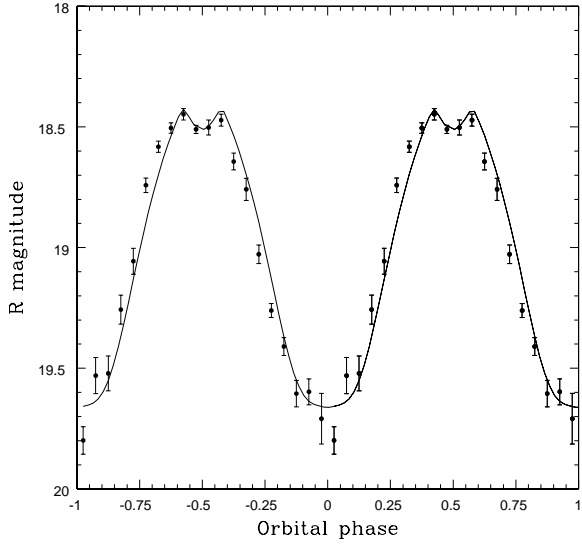}} 
\noindent 
\end{picture}

\caption[{\bf Figure 5}] {Decay light curve (16 Aug) (points) and model
fits(solid line) for L$_{x}$=1.3$\times$10$^{36}$erg~s$^{-1}$, i=72$^{\circ}$, 
q=4.2 (M$_{2}$=0.6M$_{\odot}$), R$_{disc}$=0.6 R$_{\rm L_1}$ and 
$\alpha$=5$^{\circ}$.7 
\label{fig5}}
\end{center} 

\end{figure} 

\begin{figure}
\begin{center}
\begin{picture}(150,230) 
\put(0,0){\includegraphics{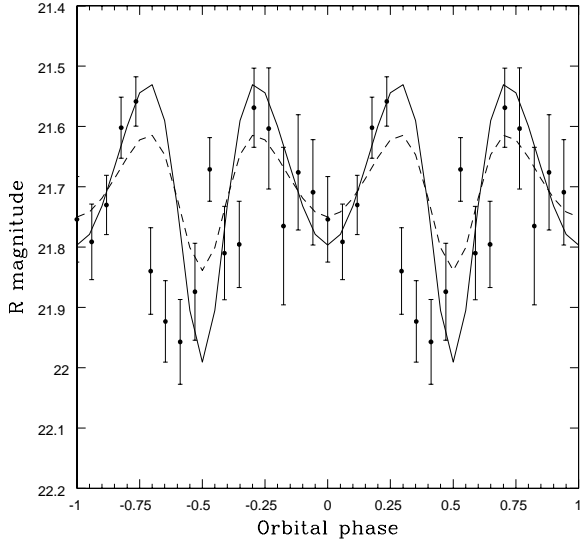}} 
\noindent 

\end{picture}

\caption[{\bf Figure 6}] {Quiescent light curve (26--27 Aug 98 and 24--26
Sep98) (points) and model plot (solid line) for i=73$^{\circ}$, 
q=4.6 and no X--ray heating. The dashed line 
shows a model plot assuming 50 per cent contamination by the accretion disc.  
\label{fig6}}
\end{center} 
\end{figure} 


\end{document}